\documentclass{aa}
\usepackage{epsf}

\newcommand{\rin}{R_{\rm in}}
\newcommand{\rout}{R_{\rm out}}
\newcommand{\ledd}{L_{\rm Edd}}
\newcommand{\lesssim}{\mathrel{\hbox{\rlap{\hbox{\lower4pt\hbox{$\sim$}}}\hbox{$<$}}}}
\newcommand{\gtrsim}{\mathrel{\hbox{\rlap{\hbox{\lower4pt\hbox{$\sim$}}}\hbox{$>$}}}}

\begin{document}

\thesaurus{08(02.01.2; 02.16.2; 02.18.7; 08.02.1; 08.14.1; 13.25.1)}

\title{Scattering in the inner accretion disk and the waveforms and
polarization of millisecond flux oscillations in LMXBs}

\author{S.Y. Sazonov\inst{1,2}
\and R.A. Sunyaev\inst{2,1}}

\offprints{S. Sazonov (ss@hea.iki.rssi.ru)}

\institute{Space Research Institute (IKI), Profsouznaya~84/32, 117810
Moscow, Russia
\and MPI f\"ur Astrophysik, Karl-Schwarzschild-Str.~1, 86740 Garching bei
M\"unchen, Germany}

\date{Received; accepted}

\titlerunning{Scattering in the disk and millisecond oscillation waveforms}

\authorrunning{Sazonov \& Sunyaev}

\maketitle

\begin{abstract}
The scattering by the inner accretion disk of X-ray radiation generated near
the surface of a spinning neutron star in a low-mass X-ray binary
(LMXB) has observable effects on the waveforms of millisecond X-ray
flux oscillations produced e.g. during type-I bursts or in the
millisecond pulsar SAX J1808.4--3658. We study these effects in the
framework of a simplified model in which there is a single emitting spot on
the stellar surface, which is visible both directly and in X-rays
scattered from the disk. The main signature of scattering from a thin
disk is that the pulse of scattered flux leads (if the star rotates in the 
same sense as the disk) or lags (in the contrary case) the primary
pulse of direct emission by a quarter of a spin cycle. This is caused
by Doppler boosting of radiation in the sub-relativistic Keplerian
flow. The disk-scattered flux is revealed better in energy-resolved
waveforms and the phase dependence of the polarized flux
component. The phenomenon discussed permits direct testing of the
presence of standard thin disks near the neutron stars in LMXBs and 
should be observable with future X-ray timing experiments having a few
times better sensitivity than RXTE and also with sensitive X-ray polarimeters.


\end{abstract}

\section{Introduction}

The inner regions of the accretion disk around a weakly magnetic neutron
star in a low-mass X-ray binary (LMXB) system intercept and re-emit a
significant fraction of radiation generated near the surface of the neutron
star. About half of the X-ray luminosity liberated via accretion
originates in a boundary layer of the disk or in a spreading layer on
the stellar surface. During type-I X-ray bursts, the source luminosity
almost entirely originates on the stellar surface. Therefore,
scattering of the central emission in the inner disk should always
have an important effect on the properties of the observed X-ray
signal (Lapidus \& Sunyaev 1985).

\begin{figure*}
\epsfxsize=\hsize
\epsffile{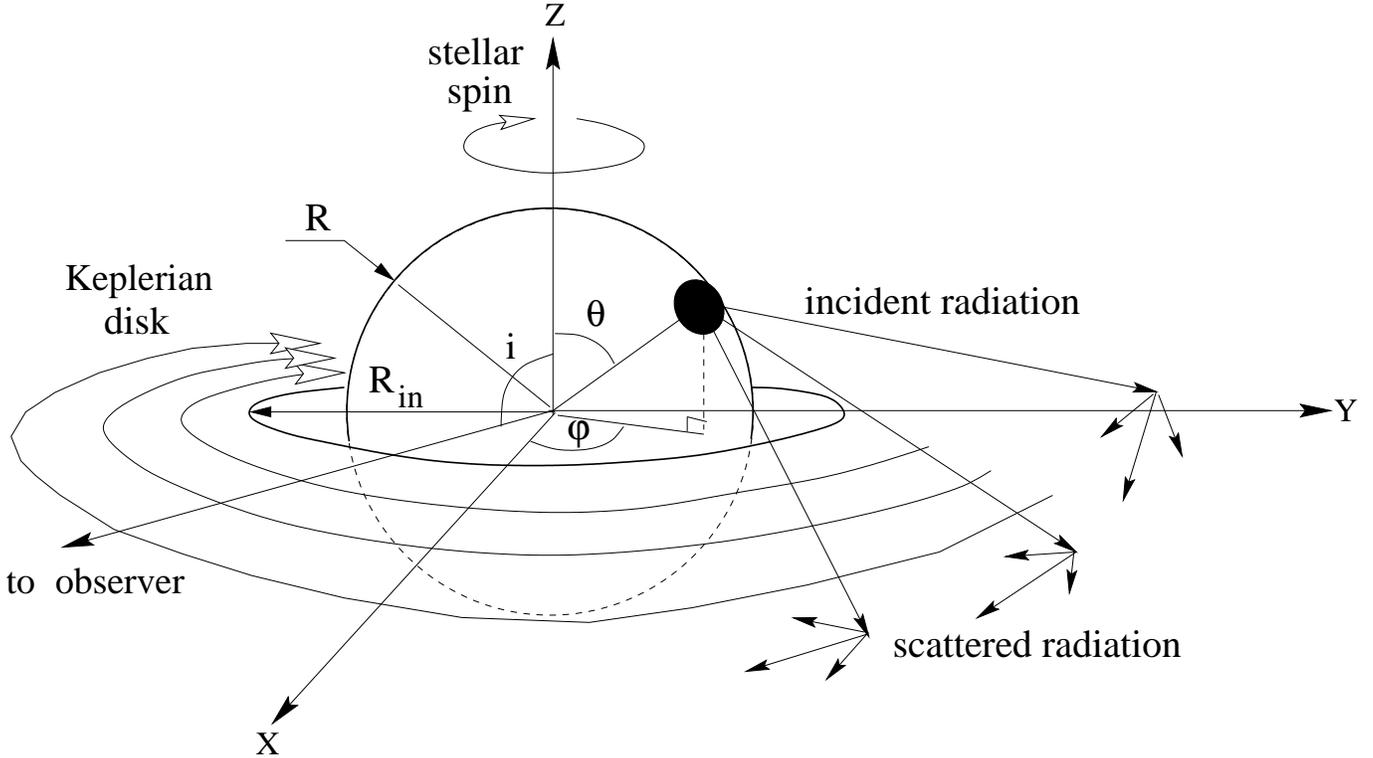}
\caption{Schematic picture of scattering of radiation emitted by 
a spot on the surface of a spinning neutron star from a surrounding Keplerian
accretion disk.
}
\label{sketch}
\end{figure*}

The Rossi X-ray Timing Explorer (RXTE) detections of millisecond X-ray flux
oscillations from many sources have demonstrated that the neutron stars in
LMXBs are rapidly spinning, with frequencies ranging between 300 and 600~Hz
(see Strohmayer 2000; Van der Klis 2000 for reviews). These
oscillations are likely the result of azimuthal asymmetry in the
generation of X-rays, e.g., there is a single emitting spot on the
star (see Fig.~\ref{sketch}). In future X-ray timing experiments, a
wealth of information on the geometrical and physical properties of
neutron stars and the inner accretion flows in LMXBs will be obtained
from quality measurements of oscillation waveforms. These can be
non-sinusoidal for several reasons. First, if one considers the
situation (as is frequently done) in which there is {\sl no accretion
disk} near the neutron star, then only half (neglecting aberration of
light) of the surface of the spinning star can be visible an any
instant. Therefore, periodic eclipses of the emitting zone may
take place. Second, since the spin velocity at the stellar equator is of
the order of $0.1 c$ (the speed of light), the emission pulse should
be asymmetric due to Doppler boosting (the spot repeatedly moves
towards and away from us --- see Fig.~1; Chen \& Shaham 1989; Weinberg
et al. 2000). Both general relativistic light bending and finite spot
size act to reduce oscillation amplitudes (Pechenick et al. 1983; Wood
et al. 1988; Riffert \& Meszaros et al. 1988; Weinberg et
al. 2000). 

If we take into account the accretion disk lying near the neutron star, it
will still be possible for the above phase-dependent effects to take
place, but additional effects will appear at the same time. In this
paper we study the {\sl role of the accretion disk} in forming
oscillation waveforms. More specifically, we show that there are clear
observational effects that allow us to test the presence of a
standard thin disk near neutron stars in LMXBs. First, such a disk
screens half of the star from our view. Second, if the emitting zone
is located above the disk (with respect to us), part of its flux is
scattered by the disk in our line of sight.

\begin{figure*}
\epsfxsize=\hsize
\epsffile[0 160 590 700]{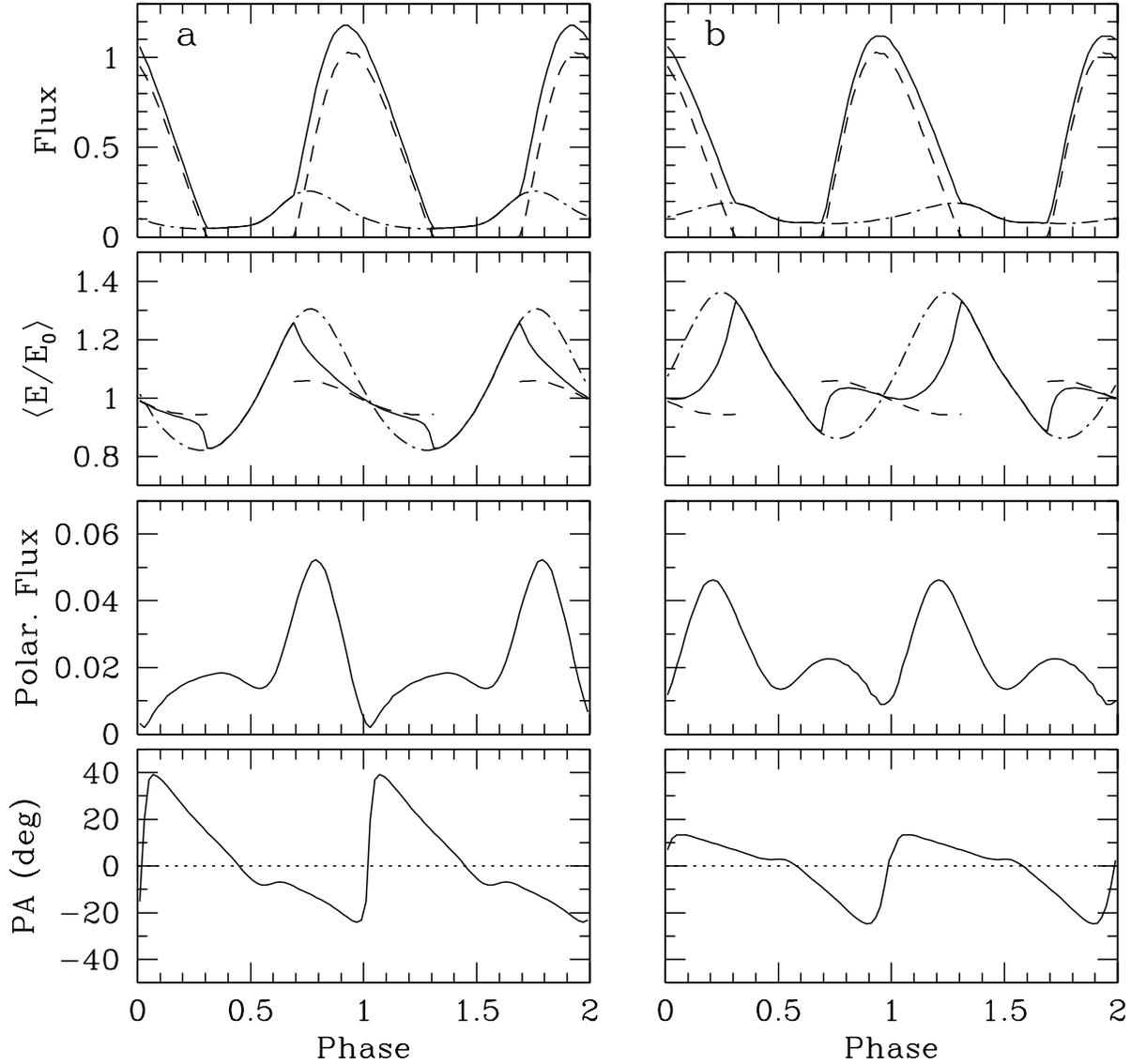}
\caption{(a) Spin-phase dependences of various characteristics of emission
from a LMXB, taking into account the effect of disk-scattering, for the
basic model ($R=\rin=6GM/c^2$), $i=60^\circ$, 
$\theta=60^\circ$, a spin frequency of $f=300$~Hz, a point-like emitting
spot ($\alpha=0$) and corotation of the star and disk. The dashed curves
correspond to direct emission from the spot, the dash-dotted lines to the
disk-scattered radiation, and the solid lines to the sum of these two
components. The panels from top to bottom are for: photon flux, mean photon
energy increment, polarized flux (in the same units as total flux) and
polarization angle. Note that only the scattered component contributes to
the polarized flux. (b) Same as (a), but for the case of star-disk
counterrotation.
}
\label{profile}
\end{figure*}

In the innermost regions of the disk, matter is rotating with speeds
$\sim 0.5$~c. Therefore, scattered photons are Doppler-boosted along
the direction of the disk rotation: i.e., towards us when the bright
spot is irradiating the sector of the disk in which matter approaches
us (which approximately corresponds to the situation presented in
Fig.~\ref{sketch}), and away from us when the opposite sector of the
disk is being irradiated (half a cycle later). As a result, the phase
dependence of the scattered flux component will be distinctly
different from that of the direct emission of the spot. Namely, the 
pulse of scattered emission will {\sl lead} the pulse of direct 
flux by roughly a quarter of a cycle. Furthermore, X-ray flux registered
during this secondary pulse will be substantially harder and more polarized
(in absolute units) than at other phases (Fig.~\ref{profile}a). These
conclusions only apply to binaries in which the neutron star rotates in the
same sense as the disk. If the star rotates in the opposite sense
(Fig.~\ref{profile}b), the scattered component will {\sl lag} the direct
signal, by the same quarter of a cycle. This makes it possible to uncover
counterrotating LMXBs (see Sibgatullin \& Sunyaev 2000 on possibilities of
formation of such systems).

General relativity predicts the existence of a gap between the last stable
Keplerian orbit and the stellar surface for some (soft) neutron star
equations of state (see, e.g., Kluzniak \& Wilson 1991). In geometries with
a gap, part of the lower stellar hemisphere is visible, and also less flux
from the upper hemisphere is scattered in the disk. However, recent
calculations (Sibgatullin \& Sunyaev 2000) show that there will be no gap if
the neutron star rotates with frequency $\sim 600$~Hz in the same sense as
the disk. For the slower rotation at 300~Hz, a thin gap can exist, but its
width proves to be smaller than the height of the boundary/spreading layer
for the accretion rates typical of LMXBs (Inogamov \& Sunyaev 1999; Popham
\& Sunyaev 2000). On the other hand, a large gap can arise in the case of
counterrotation. We discuss some consequences of the presence of a gap in \S
3.2.

A similar situation probably takes place in the millisecond pulsar SAX
J1808.4--3658, a unique object that demonstrates coherent 2.5~ms flux
oscillations (Wijnands \& Van der Klis 1998; Chakrabarty \& Morgan 1998;
Gilfanov et al. 1998). In this system, matter supposedly flows towards two
nearly antipodal magnetic polar caps on the neutron star, and the accretion
disk terminates at some distance from the stellar surface. Then, the hot
spot located below the disk could be observable. However, the measured
nearly sinusoidal pulse profile (Wijnands \& Van der Klis 1998) suggests
that this spot contributes very little, if any, to the total signal. This 
fact can be used for constraining the geometry and inclination of the
system. The upper hot spot should be observable both directly and in X-rays
scattered from the disk.

Our scenario may apply to type-I X-ray bursts. Indeed, RXTE detections
of nearly coherent oscillations during bursts from eight sources (see
Strohmayer 2000 for a review; Wijnands et al. 2000; Galloway et al. 2000)
imply the presence of a localized emitting zone. Apparently,
this zone slowly evolves in time (on sub-second timescales $\gg$ the spin
period), as the thermonuclear burning front propagates over the
stellar surface. It is not clear whether the innermost regions of a
standard accretion disk can survive during a burst, when the disk is
subject to strong radiation forces. In particular, radiation drag causes the
matter flowing from the disk to lose angular momentum and spiral
inward (Miller \& Lamb 1993). Obviously, the disk has a greater chance
to remain intact during the rise and decay stages of a burst, when the
emitted luminosity, $L$, is much less than the critical Eddington
luminosity, $\ledd$. Note that if the inner disk is destroyed during
the maximum of the burst, it can reconstruct very quickly at some later
moment when the burst emission has faded substantially, because the
characteristic times for all processes taking place within several
stellar radii from the star are much shorter than the typical burst
duration ($\sim 10$~s).

Spots of enhanced brightness may also arise in a spreading layer on the
surface of a neutron star (Inogamov \& Sunyaev 1999). This layer rotates
differentially, with the azimuthal velocity ranging between the Keplerian 
velocity ($\sim 0.5c$) at the equator of the star, and the stellar spin
velocity ($\sim 0.1 c$) near the edges of the layer. Under such
circumstances, long-living whirls, analogous to typhoons, Jupiter's famous
red spot, or ``cat eyes'', may regularly appear in two bright
latitudinal belts within the layer that are located in the
upper and lower hemispheres equidistantly from the equator. The
velocity shear is largest in these belts. Their location is
determined by the accretion rate, namely, they become more distant from
the equator (thus closer to the poles) with increasing accretion
rate. Also, instabilitites developing in the accretion disk can
modulate the accretion rate through the neck of the disk into the
spreading layer. In particular, magnetic-field loops originating in
the disk due to Balbus-Hawley instability may penetrate 
into the layer. All these phenomena can lead to the formation of
bright emitting zones, rotating faster than the star itself, with
quasi-frequencies up to the Keplerian one of $\sim 2$~kHz (Inogamov \&
Sunyaev 2000; Sunyaev \& Revnivtsev 2000). If a spot emits a
significant fraction of the source luminosity during several spin
cycles, quasi-periodic flux oscillations will appear. Averaged (with a
fixed period) waveforms of such incoherent oscillations will tend to
be nearly sinusoidal. However, the disk-scattered flux should still
manifest itself clearly in the power density spectrum of the total
signal (i.e., in the ratio of the powers contained in different
quasi-harmonics), as well as in the phase delays ($\lesssim 0.25$)
between waveforms measured in different energy bands and in the phase
shift ($\sim 0.25$) of the polarized flux component 
with respect to total flux.

It is important to note that the considerations above are only applicable
to atoll sources, i.e., to LMXBs with luminosities between 0.01
and 0.5 $\ledd$. In this case, the approximation of a thin accretion
disk (Shakura \& Sunyaev 1973) is appropriate. At luminosities
approaching $\ledd$, like those of Z-sources, the boundary layer
extends radially more than one stellar radius and is geometrically thick
(Popham \& Sunyaev 2000). The adjacent disk also becomes thick. Therefore,
our picture is no longer valid. Similarly in the case of $L<0.01\ledd$,
Compton cooling becomes inefficient for matter in the disk, and a
geometrically thick advection-dominated accretion flow may form (e.g., Narayan
et al. 1998).

\section{Model}
 
\subsection{Parameters and physical assumptions}

Let a neutron star of mass $M=1.4 M_{\odot}$ and radius $R$ spin
with frequency $f$ and be surrounded by an infinitely thin, flat accretion
disk with inner radius $\rin$. As our {\sl basic model}, we adopt the
geometry in which $R=\rin=R_0=6GM/c^2=12.4$~km, i.e., the neutron star is
the same size as the innermost Keplerian orbit that corresponds to the
Schwarzschild metric. 

The spin axis is normal to the accretion plane and makes an angle $i$
with the line of sight. Two types of stellar spin are considered: one in
the same sense as the Keplerian flow (corotation) and the other in the
opposite sense (counterrotation).

A single homogeneously bright spot, of radius $\alpha$, on the star emits
unpolarized radiation in accordance with the Lambert law in its rest
frame, i.e. $F(\mu)\sim\mu$ (where $\mu$ is the cosine of the angle
between the normal to the spot and the viewing direction). The spot is
centered at an angular distance $\theta$ from the pole located in the
upper hemisphere (see Fig.~\ref{sketch}).

We also performed several computational runs, without treatment of
polarization effects, for a more realistic situation in which the spot
emits as an electron-scattering atmoshpere, i.e.,
$F(\mu)\sim 3\mu(1+2\mu)/7$. In this case, radiation emerges linearly 
polarized from the spot and is more beamed. We found that the effect
of the additional beaming on the phase dependence of flux (including
the disk-scattered component) is fairly small and likely of the same
order as the neglected effect of gravitational light bending (see
below). The intrinsic polarization of the radiation emitted by the spot is
also small for typical values of the angles $i$ and $\theta$,
as follows directly from the classical results (Sobolev 1949; Chandrasekhar
1950). Indeed, the $\mu$--dependence of the degree of polarization,
$P(\mu)$, has a sharp maximum of 11.7\% at $\mu=1$. This maximum
corresponds to a spot position at the limb of the star, when no flux
is detectable. The polarized flux from the spot, i.e., $P(\mu)F(\mu)$,
peaks in the course of stellar rotation when $\mu\sim 0.5$, and the
maximum polarized flux, $F_{max}^{\rm pol}$, is typically only $\sim
1$\% of the maximum total flux, $F_{max}$. The phase-dependent
polarization effect due to disk-scattering proves to be much more
significant (see Fig.~\ref{profile}), namely 
$F_{max}^{\rm pol}/F_{max}\sim  5$\% (for the sum of direct and
scattered radiation) for typical values of $i$ and $\theta$. Based on
these arguments, we restrict ourselves to considering below only the
case of a spot emitting as a Lambert source.

Matter in the disk is in perfect circular Keplerian rotation; the
velocity field (dependence on the radial coordinate $r$) is calculated as
appropriate for the Schwarzschild metric:
\begin{equation}
v_\phi(r)=\left[2\left(\frac{3r}{R_0}-1\right)\right]^{-1/2} c,
\end{equation}
so that $v_\phi(R_0)=0.5 c$. 
  
The disk is perfectly reflective, which is a reasonable approximation
for its inner region. In fact, even if H-like and He-like ions of iron
are abundant in the inner disk, these can only effectively absorb
photons with energies lying in a narrow band above 9~keV. For softer
X-ray radiation, the reflection albedo of the inner disk should be
almost unity. Note also that the disk's surface layer having a Thomson
depth $\tau\gtrsim 1$ is probably completely ionized due to
irradiation by X-rays from the neutron star (e.g., Basko et al. 1974;
Nayakshin et al. 2000). For the above reasons, 
the only radiation process we consider is Thomson scattering. Photons incident
on the disk typically get scattered away after a few ($N<10$) scatterings.
With the temperature of matter in the disk $kT\sim 1$~keV (Shakura \&
Sunyaev 1973) and for incident radiation with $E\lesssim 10$~keV, the photon
energy changes on average by a small fractional amount during the multiple
scatterings: $|\Delta E/E|\sim |N (4kT-E)/mc^2|< 0.1$ [of course, the
spectrum is also Doppler-broadened: $(\Delta E/E)^2\sim 2N kT/mc^2$]. For
this reason, we consider scatterings to be coherent in the rest frame of
the Keplerian flow. To calculate the angular distribution and polarization
state of scattered radiation (Thomson scattering produces linear
polarization) as measured in the disk rest frame, we apply the classical
formulae of the theory of radiation transfer in plane-parallel atmospheres
for the Rayleigh angular scattering law (Chandrasekhar 1950).

General relativistic light bending is not taken into account. As a
result, photon trajectories are straight lines, which greatly
simplifies the computations. At the same time, this simplification cannot 
affect the main results of our study qualitatively. We do not
consider the redshifting of radiation either, because its magnitude is the
same (for the Schwarzschild geometry) for all photons, due to their common
origin on the stellar surface.

All possible shadowing effects are taken into account. For the direct
emission, these include: (1) the invisibility of the bright spot when it is
turned away from the observer (provided that $i+\theta> 90^\circ$), and
(2) the screening by the disk ($\theta>90^\circ$). The neutron star will, in
turn, intercept part of the radiation scattered by the disk in the
observer's direction if $i+\theta> 90^\circ$ (the just quoted constraints
are appropriate for the basic model with a point-like spot). Note that our
model overestimates the duration of any eclipse intervals, due to
the neglect of the light bending effect.
 
\subsection{The calculation procedure}

We introduce a reference frame, static with respect to the observer,
with origin at the center of the neutron star (see Fig.~\ref{sketch}). The
accretion disk lies in the XY-plane, and the Z-axis is the spin axis of the
star. For a given inclination angle $i$, the observer is at infinity in the
direction ${\bf l}=(\sin i,0,\cos i)$. Instantaneous positions of the
center of the emitting spot are defined by the azimuthal angle $\phi$, in
addition to the previously defined parameter angle $\theta$. 

The calculation essentially consists of following, both in space and time,
the tracks of a large number of photons emitted by the bright spot.
Multi-dimensional integration is carried out over $\phi$ (from 0 to $2\pi$),
as well as over many zones that compose the total area of the bright spot
and the disk. The integration over the disk is limited radially by an
outer boundary of $\rout=11 R_0$. Only a small fraction, $< 10$\%, of the total
X-ray flux illuminating the whole disk is intercepted by yet farther
regions (if the warping of the disk is neglected), causing their role in
forming the scattered signal to be practically unimportant. It is also
worth mentioning that for $\theta>60^\circ$ (which is expected to occur with
$1/2$ probability if spot positions are random), more than 75\% of the
radiation incident on the disk is received by the innermost region
of $R_0<r<4 R_0$.

On output, we obtain the arrival time, energy shift ($E/E_0$), and polarization
state for each tested photon. Energy shifts have a Doppler origin:
either due to the stellar spin or due to the Keplerian rotation of the
disk. Our code  calculates the direct signal, the component scattered
from the disk and the sum of these two. Several types of final results
can be produced: (1) oscillation waveforms, i.e., the number of 
photons as a function of spin phase $\xi$ ($\xi=$ arrival time/spin period),
(2) the phase dependence of the average energy increment, $\langle
E/E_0\rangle(\xi)$ (provided that the spot emits monochromatic
radiation of energy $E_0$), (3) oscillation waveforms in several
specified narrow energy bands for a given energy spectrum of the spot
emission, and (4) the phase dependence of the polarized flux component
(namely, amplitude and the position angle of the polarization plane).

The zero of $\xi$ is defined as follows. Suppose the bright spot is at
some moment in the position $\phi=0$ and emits a photon in the
direction ${\bf l}$ (along the line of sight). Imagine that at the same
instant another photon is emitted from the center of the star, ${\bf
r}=(0,0,0)$, also in the direction ${\bf l}$. We postulate that the second
photon will arrive at the detector at phase 0. This implies that the first
photon will reach the detector a little earlier, at
$\xi=1-Rfc^{-1}\cos(\theta-i)$. Note that this phase difference, which has a
time-of-flight origin, is small for $f< 600$~Hz: $1-\xi< 0.025$. Larger
phase delays of this type can arise for photons that experience a
disk-scattering on their way to the observer. The Thomson-scattering process
itself is assumed to introduce no extra phase delay, which is true for our
case of a geometrically thin disk.

One phenomenon is worth mentioning here. As the neutron star spins, the
zone of the disk that is illuminated by the bright spot rotates with a 
speed of $2\pi f r$, which exceeds the speed of light far enough from
the star. For example, at $f=600$~Hz, this occurs at $r>80$~km. Such distant
regions, however, intercept very little of the spot emission. For this
reason, the contribution of superluminal effects to the phase dependence of
the scattered component is negligible.

Let us now turn to the main physical mechanism at work, namely, Doppler
boosting of radiation, which operates both during emission from the
spinning star and during scattering from the disk. The disk-related effect
is the larger, as follows from a comparison of the Keplerian velocity, $\sim
0.5c$, with the spin velocity at the stellar equator, $0.078(f/300\,\,{\rm
Hz})c$. Our code makes use of a set of equations giving the Lorentz
transformations of photon energy $E$, direction of propagation ${\bf k}$,
solid angle $d\Omega$, and the electric (${\bf E}$) and magnetic (${\bf H}$)
field vectors (Landau \& Lifshits 1975):
\begin{equation}
E^\prime = E\gamma\left(1-\frac{{\bf k}{\bf v}}{c}\right),
\label{lorentz_1}
\end{equation}
\begin{equation}
{\bf k^\prime} = \frac{1}{\gamma(1-{\bf k}{\bf v}/c)}\left\{
{\bf k}+\left[\frac{c(\gamma-1)}{v^2}{\bf k}{\bf v}-\gamma\right]
\frac{{\bf v}}{c}\right\},
\label{lorentz_2}
\end{equation}
\begin{equation}
d\Omega^\prime = \frac{1}{\gamma^2 (1-{\bf k}{\bf v}/c)^2}\,d\Omega,
\label{lorentz_3}
\end{equation}
\begin{eqnarray}
{\bf E}_\parallel^\prime &=& {\bf E}_\parallel,\,\,\, 
{\bf E}_\perp^\prime=\gamma\left({\bf E}+\frac{[{\bf v}{\bf H}]}{c}
\right)_\perp,
\nonumber\\
{\bf H}_\parallel^\prime &=& {\bf H}_\parallel,\,\,\,
{\bf H}_\perp^\prime=\gamma\left({\bf H}-\frac{[{\bf v}{\bf E}]}{c}
\right)_\perp,
\label{lorentz_4}
\end{eqnarray}
\[
{\rm where}\,\,\, \gamma=\left[1-\frac{v^2}{c^2}\right]^{-1/2}.
\]
Here, the unprimed terms are measured in the observer's frame,
whereas the primed ones in a moving (with velocity $\bf v$) frame, which
can be either the rest frame of the bright spot or the rest frame of the
Keplerian flow; the subscripts $\parallel$ and $\perp$ denote a vector's
components along and normal to ${\bf v}$, respectively. Equations
(\ref{lorentz_4}) govern the Lorentz transformation of the polarization
plane. Note that the polarization degree is a Lorentz invariant (e.g., Landau
\& Lifshitz 1975).
 
An individual disc-scattering act is computed in three steps. At first,
the energy, $E_0$, the direction of propagation, ${\bf k_0}$, and the photon
flux of the incident unpolarized radiation are transformed to the
rest frame of the Keplerian flow, using
Eqs.~(\ref{lorentz_1})--(\ref{lorentz_3}). Then, the photon flux emergent in 
the direction ${\bf l^\prime}$ and its polarization state are
calculated in this frame, following Chandrasekhar (1950). Finally,
the reverse transition to our frame is implemented, using
Eqs.~(\ref{lorentz_1})--(\ref{lorentz_4}). The emergent photon energy
can also be directly related to the incident energy:
\begin{equation}
E=E_0\frac{1-{\bf k_0}{\bf v}/c}{1-{\bf k}{\bf v}/c}.
\end{equation}

\section{Results}
\subsection{Point-like emitting spot}

We begin by considering the situation in which the bright spot is infinitely
small. Most of our results are obtained for the basic model ($R=\rin=R_0$) and
$f=300$~Hz.

In Fig.~\ref{profile}, we have plotted the phase dependences, which
correspond to $i=60^\circ$ and $\theta=60^\circ$, of photon flux, mean
energy increment (for an initially monochromatic spectral line), polarized
photon flux and polarization position angle (which is measured relative to the
meridional plane --- defined by the normal to the disk and the line of
sight, and takes values between $-90^\circ$ and $90^\circ$). The cases
of star-disk corotation and counterrotation are compared. 

As expected, the pulse of direct emission is asymmetric, in particular, flux
peaks at $\xi\approx 0.95$, rather than at $\xi=1$, which would be the case
if the star were spinning very slowly. The maximum of $\langle E/E_0\rangle$
for the direct signal is at $\xi\approx 0.75$, which corresponds to the spot
approaching us. These spin-induced Doppler effects are well-known (see,
e.g., Weinberg et al. 2000). The spot is not visible directly during the
eclipse centered at $\xi=0.5$. The direct emission is unpolarized
by assumption. 

As regards the scattered component, flux, mean energy increment and
polarized flux all peak near $\xi=0.75$ (i.e., before the peak of direct
emission) in the case of corotation, but near $\xi=0.25$ (i.e., after 
the main peak) in the case of counterrotation. The partial obscuration of the
disk by the star, which occurs around $\xi=0.5$, plays a minor role for the
given values of $i$ and $\theta$, but becomes more important for larger
values of these angles. It is obvious from the phase dependences for the
total signal (the sum of the direct and scattered components) shown in
Fig.~\ref{profile} that the presence of disk-scattered flux could be 
most clearly established by means of X-ray polarimetry.

\begin{figure}
\epsfxsize=\hsize
\epsffile{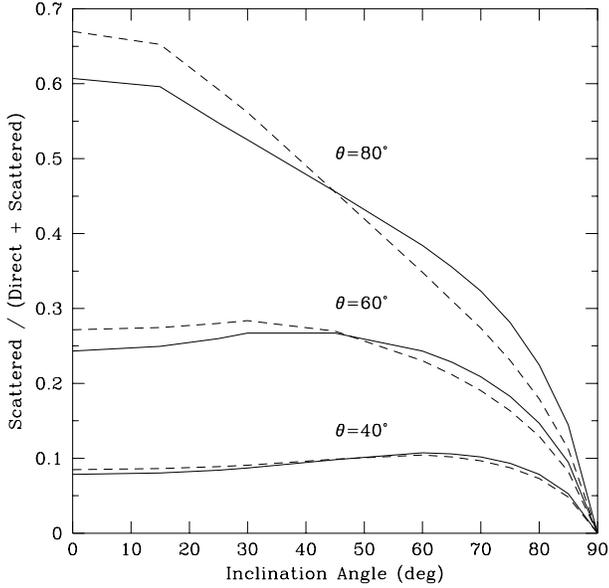}
\caption{Relative contribution of the disk-scattered component to the 
phase-integrated photon flux as a function of the angles $i$ and
$\theta$, for the basic model with $f=300$~Hz, $\alpha=0$ and the case of
corotation (solid lines). For comparison, the results for an imaginary
non-rotating disk and very slowly spinning neutron star ($f=3$~Hz) are
also shown (dashed lines).
}
\label{flux}
\end{figure}

\begin{figure}
\epsfxsize=\hsize
\epsffile{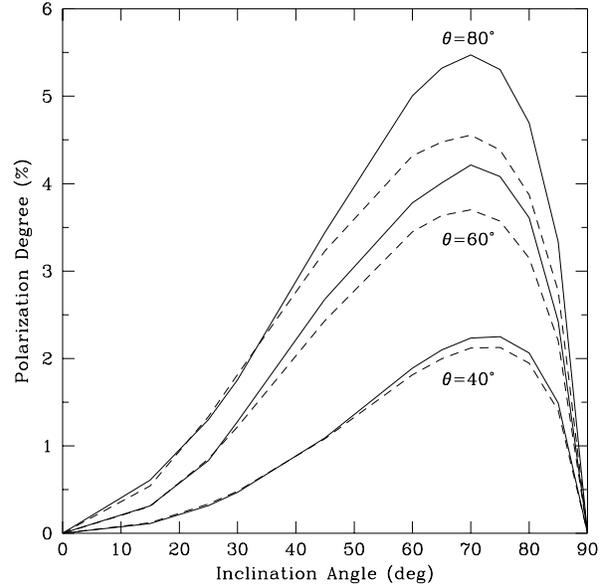}
\caption{Polarization degree of the phase-integrated total (direct
plus scattered) flux as a function of the angles $i$ and $\theta$, for
the basic model with $f=300$~Hz, $\alpha=0$ and the case of corotation
(solid lines). For comparison, the results for an imaginary
non-rotating disk and very slowly spinning neutron star ($f=3$~Hz) are
also shown (dashed lines).
}
\label{pol}
\end{figure}

The amplitude of the disk-scattering feature in an oscillation waveform is
proportional to the contribution of scattered flux to the time-integrated
signal, which can be found by integrating the phase dependence. In
Fig.~\ref{flux}, the relative contribution of scattered flux is 
shown as a function of the inclination angle for three values of $\theta$,
and $f=300$. These dependences are compared with the dependences that would
result if the disk were non-rotating and the neutron star were spinning very
slowly ($f\rightarrow 0$), so that no Doppler boosting of radiation would
take place. One can see that the differences between the two sets of curves 
are small, i.e., Keplerian motion has a much smaller effect on the integral
of flux than on its phase dependence. The spin frequency proves to be
yet less important here. We can draw the following conclusions 
from Fig.~\ref{flux}. The relative contribution of the scattered component is
less than 10\% for $\theta<40^\circ$. At lower positions (with respect to the
disk) of the bright spot, this contribution can be much larger (up to 100\%
at $i=0$$, \theta=90^\circ$). Provided that bright spot positions are
random, $60^\circ<\theta<90^\circ$ in 50\% of situations. Similarly,
$60^\circ<i<90^\circ$ for every second binary. Therefore, scattered emission
should typically contribute $\sim 20$\% to the total signal.

Fig.~\ref{pol} is similar to Fig.~\ref{flux}, but for the polarization
degree of the phase-integrated total flux. Typically, the polarization
degree is $\sim 3$\% and is somewhat enhanced owing to the Doppler boosting of
radiation in disk-scattering. Our results for a non-rotating
disk and slowly spinning neutron star presented in Fig.~\ref{flux} and
Fig.~\ref{pol} are in agreement with the results of Lapidus \& Sunyaev (1985).

\begin{figure}
\epsfxsize=\hsize
\epsffile{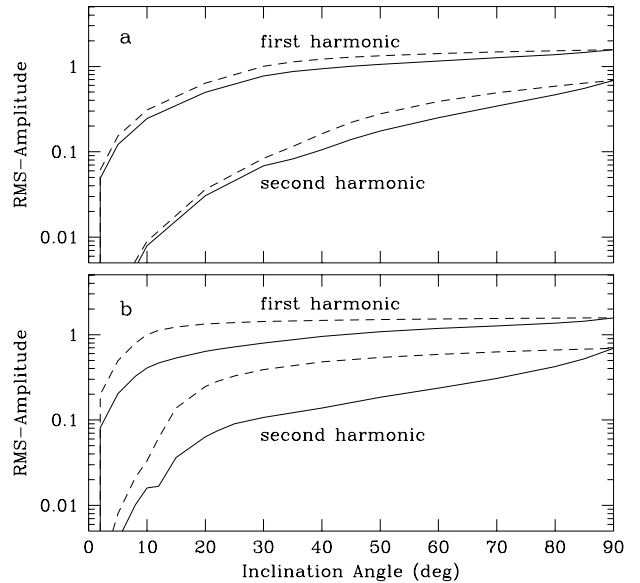}
\caption{Amplitudes of the first and second Fourier harmonics in the
direct flux from the spot (dashed lines) and in the total signal (solid
lines) as a function of the inclination angle $i$, for the basic model with
$f=300$~Hz, $\alpha=0$ and the case of corotation. (a) $\theta=60^\circ$,
(b) $\theta=80^\circ$.
}
\label{harmon}
\end{figure}

Let us now examine the effect of the scattered component on the harmonic
content of flux oscillations. To this end, oscillation waveforms need to be
Fourier-transformed. In Fig.~\ref{harmon}, we compare the rms-amplitudes of
variability contained in the first two harmonics. Note that much of the
harmonic content of the direct emission is due to the Doppler boosting by the
spinning star; more details on this can be found in (Weinberg et al. 2000).
The scattered component generally causes the total signal to be more
sinusoidal (i.e., the ratio of the amplitude of the second harmonic to that
of the first harmonic decreases). This effect is mostly the result of the
presence of scattered flux during eclipses of the bright spot (see
Fig.~\ref{profile}) and becomes more pronounced for lower spot positions 
(see Fig.~\ref{harmon}b).

\subsection{Effects of the disk/star size ratio, spin frequency and
spot size}

\begin{figure}
\epsfxsize=\hsize
\epsffile[0 380 580 720]{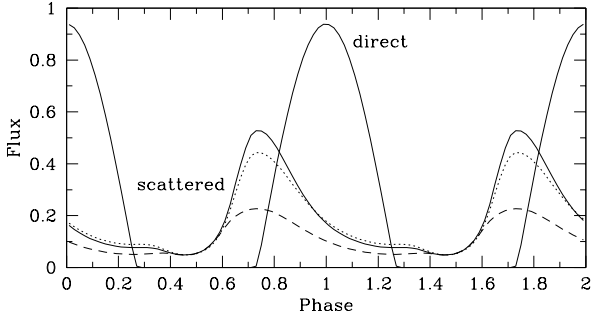}
\caption{Effect of the ratio of the disk inner radius, $\rin$, to the
stellar radius, $R$, on the oscillation waveform of the disk-scattered
component, for $i=60^\circ$, $\theta=80^\circ$, $\alpha=0$ and the case of
corotation. A low spin frequency, $f=3$~Hz, is taken, so that the Doppler
boosting by the star is unimportant, and the direct signal (the solid line
with the corresponding label) is independent of $R$. Three geometries are
considered: (1) the basic model ($R=\rin=6 GM/c^2$; solid line), (2)
$R=\rin=8 GM/c^2$ (dotted line) and (3) $R=6 GM/c^2$, $\rin=8 GM/c^2$
(dashed line).
}
\label{rad}
\end{figure}

\begin{figure}
\epsfxsize=\hsize
\epsffile[0 380 580 720]{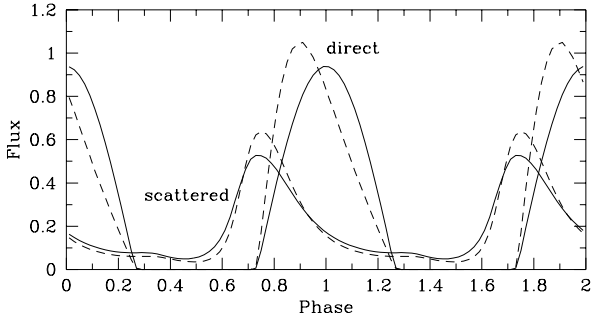}
\caption{Effect of the spin frequency, $f$, on the oscillation waveforms of
the direct and scattered flux components, for the basic model with
$i=60^\circ$, $\theta=80^\circ$, $\alpha=0$ and the case of corotation. Two
cases are considered: one of very slow rotation ($f=3$~Hz; solid lines), and
the other of fast rotation ($f=600$~Hz; dashed lines).
}
\label{freq}
\end{figure}

How crucial is the condition $R=\rin=R_0$, which defines our basic
model? In Fig.~\ref{rad}, three different oscillation waveforms of
disk-scattered radiation are shown, all corresponding to $i=60^\circ$ and
$\theta=80^\circ$. In one case, the geometry is standard. In the second
case, the neutron star extends beyond the last stable Keplerian orbit:
$R=\rin>R_0$. In the third case --- $R=R_0<\rin$, which may approximately
correspond to the millisecond pulsar SAX~J1808.4--3658, a transparent zone
exists between $R$ and $\rin$. One can see that the amplitude of the scattered
component is smaller by a factor of 2 in the case of this gap geometry. The
waveforms for the ``normal-size'' and ``oversize'' neutron stars are only
slightly different. We conclude that the oscillation phase dependence and
amplitude are almost independent of the stellar size as long as
$R=\rin\gtrsim R_0$. On the other hand, the scattered component is sensitive
to the ratio $\rin/R$, particularly for near-equatorial spot positions, if a
gap separates the star from the accretion disk.

Another parameter that can have an effect on oscillation waveforms is the
spin frequency, $f$. In Fig.~\ref{freq}, we compare two cases: one of very
slow rotation, $f=3$~Hz, and the other of fast rotation, $f=600$~Hz. In the
first case, there is no Doppler boosting of the direct emission. In the
second case, when the equatorial spin velocity is $0.16c$, the resulting
asymmetry is very pronounced, as the direct flux peaks at $\xi\approx 0.9$.
However, the scattered component is relatively insensitive to $f$.

\begin{figure}
\epsfxsize=\hsize
\epsffile{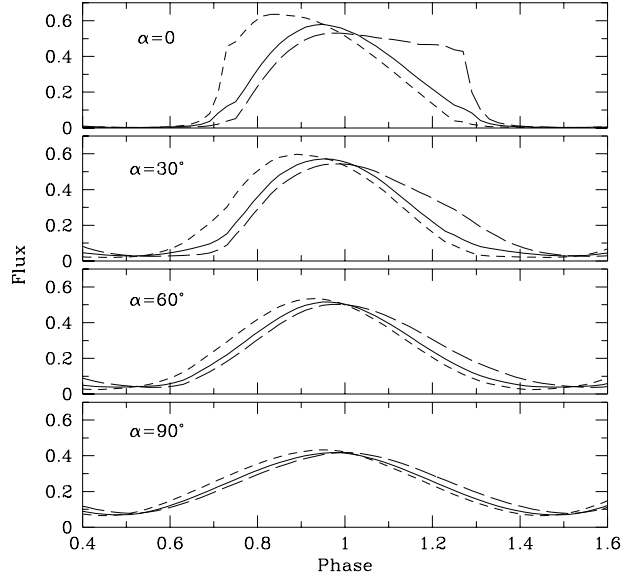}
\caption{Effect of the bright spot size on the oscillation waveform of total
(direct plus scattered) flux, for the basic model with $i=60^\circ$,
$\theta=90^\circ$ and $f=300$~Hz. Three cases are compared: of corotation
(short-dashed lines), of counterrotation (long-dashed lines) and also of a
non-rotating disk (solid lines). Note the different than in the other figures
phase region, which permits a better view of the waveforms. 
}
\label{size}
\end{figure}

\begin{figure}
\epsfxsize=\hsize
\epsffile{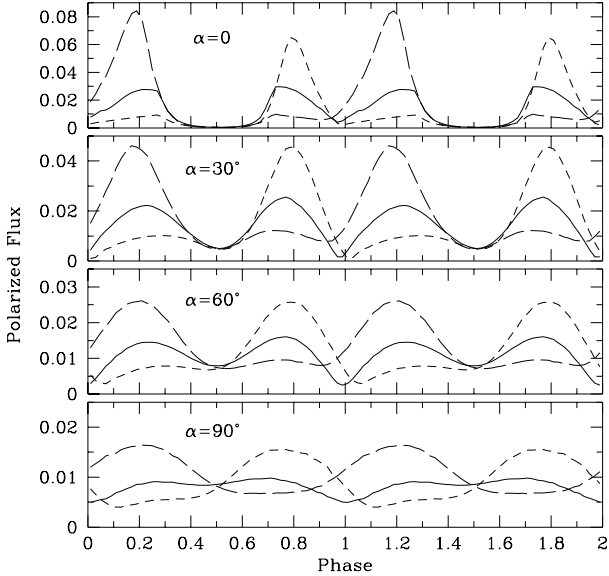}
\caption{Same as Fig.~\ref{size}, but for the polarized flux component.
}
\label{size_pol}
\end{figure}

We have so far assumed that the bright spot is point-like. In reality,
radiation is generated over some extended region of the stellar surface. Let
us examine the effect of the spot radius, $\alpha$, on oscillation
waveforms. An example is presented in Fig.~\ref{size}. A circular spot is
centered at the equator, so that half of its area is above the disk,
producing the direct and scattered flux components, while the lower half is
invisible and has no effect on the phase dependence. As expected,
waveforms become smoother with increasing $\alpha$. The asymmetry between
the rising and declining branches of the pulse due to disk-scattering becomes
less pronounced as $\alpha$ increases. As $\alpha$ approaches $90^\circ$,
the waveform becomes nearly sinusoidal, and the role of the scattered
component effectively reduces to shifting the pulse in phase.

Fig.~\ref{size_pol} is similar to Fig.~\ref{size}, but for the polarized
flux component. From these two figures one can see that the
pulse of polarized flux is always shifted relative to the pulse of total
flux by approximately a quarter of a cycle, independent of how large the bright
spot is. Whether this phase shift is negative or positive is determined by
the sense of the stellar spin with respect to the disk rotation. 

\subsection{Energy-resolved waveforms}

\begin{figure}
\epsfxsize=\hsize
\epsffile{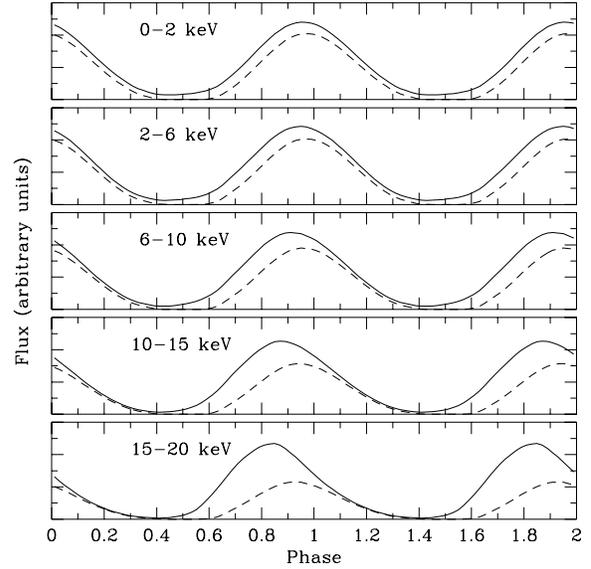}
\caption{Oscillation waveforms in five energy bands for the basic
model with $i=60^\circ$, $\theta=90^\circ$, $f=300$~Hz, $\alpha=60^\circ$,
the case of corotation and black-body spot emission with $kT=2$~keV. The solid
lines are the total signal, and the dashed lines are the direct flux component.
}
\label{spec}
\end{figure}

Photons arriving at the detector directly from the neutron
star or upon scattering in the Keplerian disk have been Doppler-shifted in 
energy. The phase-dependent fractional shift, $\Delta E/E$, 
is $\sim 0.08(f/300\,\,{\rm Hz})\cos{[2\pi(\xi+0.25)]}$ in the former case 
(for a spot at the stellar equator and an edge-on view of the
binary), and $\sim 0.5\cos{[2\pi(\xi\pm 0.25)]}$ in the latter (where
``$+$'' in the argument of $\cos$ corresponds to the case of
corotation). Therefore, energy-resolved oscillation waveforms can
provide more detailed information on the scattering of the central
emission in the disk than just the phase dependence of total flux. As
an example, we consider a situation in which a spot of radius
$\alpha=60^\circ$, centered at the equator, emits black-body radiation
with a temperature $kT=2$~keV, a value typical for type-I bursts. We
made computations for 5 energy bands covering the range
$E<20$~keV. The resulting oscillation waveforms are presented in
Fig.~\ref{spec}. The role of the scattered component apparently
becomes more important with increasing energy. Not only the relative
contribution of scattered flux becomes larger, but also the pulse is
much more asymmetric at high energies than at low.

\begin{figure}
\epsfxsize=\hsize
\epsffile{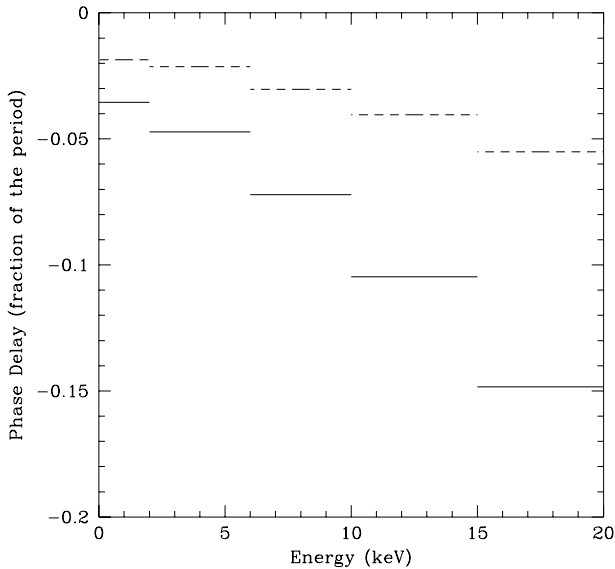}
\caption{Phase lag of the fundamental harmonic as a function of photon energy
derived from the oscillation waveforms shown in Fig.~\ref{spec}. The solid bars
correspond to the total signal, and the dashed bars to the direct flux
component.
}
\label{lag}
\end{figure}

To make the analysis more quantitative, we have Fourier-transformed the phase
dependences shown in Fig.~\ref{spec} and thus found the phase lag of the
fundamental harmonic as a function of photon energy (Fig.~\ref{lag}). We
see that the role of disk-scattering in ``delaying'' low-energy photons is
more important than the corresponding role of stellar spin. This can be
explained as follows. Both dependences plotted in Fig.~\ref{lag} approach an
asymptotic value of $-0.25$ as $E\rightarrow\infty$, because at $E\gg kT$
flux only arises at phases $\xi\sim 0.75$ (when both the spot and the 
illumination front on the disk are moving towards us). The phase lead will
constitute a significant fraction of this $0.25$ value for $E\gtrsim
kT/\langle\Delta E/E\rangle$ [because of the $\exp{(-E/kT)}$ factor in the
spectral law]. Therefore, in order to have a substantial hard-to-soft phase
difference due to disk-scattering, one needs $E\gtrsim 2kT$, as compared to 
$E\gtrsim 10(300\,\,{\rm Hz}/f) kT$ in the case of stellar rotation. Thus,
in the latter case, large phase leads ($\sim 0.1$) can only materialize very
far in the Wien tail of a black-body spectrum, i.e., in a region where
virtually no emission is generated. Note that in the case of a
counterrotating disk, hard emission will lag soft emission.

\section{Conclusion}

We have considered the effect of X-ray scattering by a thin accretion disk
on the phase dependence of spin-modulated emission from a neutron star
with a weak magnetic field. This effect makes it possible to test the
presence of standard thin accretion disks around the neutron stars in LMXBs.
The only other existing experimental method to probe Keplerian accretion
disks near compact objects (particularly black holes) is through studying
the profiles and variability of iron fluorescence lines (see Fabian et al.
2000; Reynolds 2000 for recent reviews). Furthermore, the new method
provides a unique opportunity to check whether the neutron star is
corotating with the disk, and potentially even to determine the misalignment
angle between the spin axis and the disk symmetry axis. Such measurements
can significantly enrich our knowledge of the origin and evolution of LMXBs.

The results of this paper demonstrate that the disk-scattered flux component
manifests itself better in energy-resolved oscillation
waveforms. Nearly coherent oscillations during type-I bursts,
especially those detected during the burst rise, are probably the best
candidates to look for signatures of disk-scattering. It is possible
that RXTE data can already be used for such a search. For example, the
phase lag of low-energy photons during a burst from Aql X-1 reported
by Ford (1999) may be the result of Doppler boosting
in a thin disk. It is important that this interpretation is 
self-consistent, as opposed to the alternative explanation based on Doppler
boosting by the neutron star, as shown in \S 3.3. The case of the
millisecond pulsar SAX~J1808.4--3658, for which similar phase lags and also
distortions of the pulse profile have been detected (Cui et al. 1998;
Revnivtsev 1999; Ford 2000), appears to be more complicated. The principal
difficulty here is that the observed energy spectrum has an approximately
power-law shape. If this spectral distribution originates at the hot
spot, then no energy dependence of the pulse profile can arise as 
a result of Doppler boosting (see, e.g., Chen \& Shaham 1989).

More detailed studies of the waveforms of X-ray flux oscillation in LMXBs
could be possible with $\sim 6$~m$^2$ X-ray detectors (i.e., 10 times
larger than RXTE), such as the recently proposed EXTRA and RAE missions
(Barret 2000; Kaaret 2000; http://www.cesr.fr/$\sim$barret/extra.html). A
particularly promising way to look for spin-phase signatures of
disk-scattering, at least in the millisecond pulsar SAX~J1808.4--3658,
is by means of sensitive X-ray polarimetry. The case of
SAX~J1808.4--3658 (or similar objects which may be discovered in the
future) is favourable because this source demonstrates regular pulsations
during long time intervals, which allows one to accumulate a
statistically large amount of data. On the contrary, such statistics
is hardly achievable for nearly coherent burst oscillations, or for
possible quasi-periodic oscillations originating in the spreading
layer. Previously, only the time-integrated polarization due to
Thomson scattering in the accretion disk (without allowance for its
rotation) was discussed as an observational task for near-future X-ray
polarimeters (Meszaros et al. 1988). Such polarimeter will be part of
the scientific payload of the Spectrum-X-Gamma observatory (Kaaret et al. 1993;
http://hea-www.harvard.edu/$\sim$kaaret/sxrp/index.html). Also, NASA is
currently attaching great importance to a programme of long duration balloon
flights, lasting up to 100 days (Jones 2000). This programme will no doubt
rekindle interest to balloon flights of X-ray polarimeters.

\clearpage

\end{document}